# High-Fidelity DNA Sensing by Protein Binding Fluctuations

Tsvi Tlusty,[1,2,3] Roy Bar-Ziv,[1,3] and Albert Libchaber[3]

[1]*Department of Materials and Interfaces, Weizmann Institute of Science, Rehovot, Israel 76100*
[2]*Department of Physics of Complex Systems, Weizmann Institute of Science, Rehovot, Israel 76100*
[3]*Center for Physics Biology, Rockefeller University, 1230 York Avenue, New York 10021, USA*

One of the major functions of RecA protein in the cell is to bind single-stranded DNA exposed upon damage, thereby triggering the SOS repair response. We present fluorescence anisotropy measurements at the binding onset, showing enhanced DNA length discrimination induced by adenosine triphosphate consumption. Our model explains the observed DNA length sensing as an outcome of out-of-equilibrium binding fluctuations, reminiscent of microtubule dynamic instability. The cascade architecture of the binding fluctuations is a generalization of the kinetic proofreading mechanism. Enhancement of precision by an irreversible multistage pathway is a possible design principle in the noisy biological environment.

*Kinetic proofreading and assembly fluctuations.*—Kinetic proofreading (KPR) and assembly fluctuations pertain to distinct classes of proteins. KPR is the use of energy to enhance the precision of molecular information processing [1]. For example, enzymes carrying out DNA replication reduce error rates by performing an iterative irreversible recognition process coupled to triphosphate hydrolysis [2]. Assembly fluctuations, on the other hand, are exhibited by filamentous proteins that carry out mechanical and structural functions, such as actin and tubulin [3]. In this case, triphosphate binding and hydrolysis is coupled to protein assembly or disassembly leading to collective dynamics such as treadmilling and dynamic instabilities [3,4].

In this Letter, we consider RecA, a protein that assembles into filaments, similarly to actin and tubulin. RecA filaments that form on single-stranded DNA (ssDNA) are the trigger of the SOS response to DNA damage in the cell [5,6]. We argue that RecA can "proofread" the ssDNA by its own binding fluctuations. These fluctuations are similar to microtubule dynamic instability. The assembly dynamics constitute a kinetic proofreading cascade that is a "hair-trigger" sensor of DNA length. Enhancing biomolecular precision by fluctuations, which may seem somewhat counterintuitive in a deterministic world, is presented as a natural design principle in the noisy realm of the living cell.

*RecA-DNA dynamic instability.*—The RecA-ssDNA binding kinetics has been extensively studied [5]. Assembly starts by a rate-limiting nucleation of adenosine triphosphate (ATP)–bound RecA monomer on the DNA, followed by a rapid polymerization to the 3' end of the ssDNA. Within the filament, each monomer binds to three DNA bases and hydrolyzes ATP independently. A slow, gradual disassembly, one monomer at a time, occurs at the filament 5' end when the last monomer hydrolyzes ATP [7,8]. As a result, a disassembly front vacates the DNA for recurrent nucleation-polymerization and a cascade of assembly-disassembly ensues. Previously [9], we identified this RecA assembly cascade as a simple stochastic computation process [10,11]. Here, we focus on the role of ATP-driven RecA binding fluctuations as a ssDNA length sensor, a necessity for the *in vivo* SOS response.

Our simulation of the binding kinetics exhibits an asymmetric, sawtooth pattern (Fig. 1). This is a "mirror image" of microtubule dynamic instability. Here, the almost-instantaneous nucleation is followed by slow end disassembly, whereas the microtubule dynamics is in-verted [12,13]. Strong fluctuations, of the order of the DNA length $N$, can "scan" a DNA sequence most efficiently. The time interval between nucleation events is $t_+ \sim 1/(N\kappa_{nuc}R)$, where $\kappa_{nuc}$ is the nucleation rate and $R$ is the RecA concentration. The time required to empty an average filament is of the order $t_- \sim N/2\kappa_-$. Therefore a maximal rate of large fluctuations can be achieved by matching these two time scales, $t_+ \sim t_-$. This argument suggests a dimensionless control parameter,

$$\gamma = \frac{t_-}{t_+} = \frac{\kappa_{nuc}R}{2\kappa_-} N^2 \qquad (1)$$

that tunes the fluctuations. At low $R$, $\gamma \ll 1$, the filament is almost empty and nucleation events are sparse. At saturation, $\gamma \gg 1$, nucleation events are frequent but of small amplitude and hence do not scan the entire sequence—the dynamic instability disappears. The sensitivity to DNA sequence variations (of either length or sequence) is optimal in the strong fluctuation regime, $\gamma \sim 1$, where frequent nucleation-disassembly events scan the entire sequence.

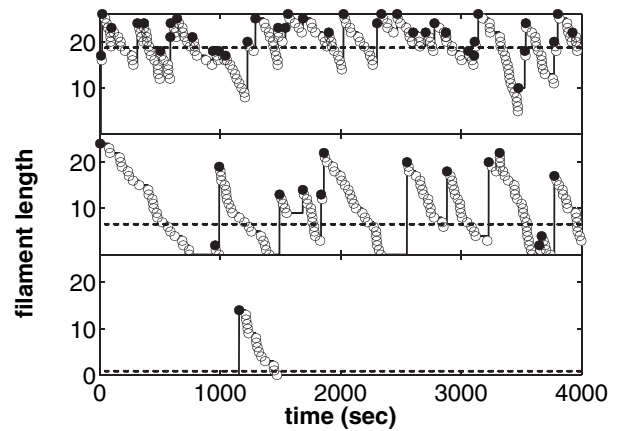

FIG. 1. Simulation of RecA binding to ssDNA with $N = 26$ binding sites, $\kappa_- = 0.05 \text{ sec}^{-1}$, and $\kappa_{\text{nuc}} = 0.006 \text{ sec}^{-1} \mu\text{M}^{-1}$. Slow disassembly (open circles) follows rapid nucleation events (solid circles). Mean binding (dashed line) and fluctuations are tuned by the control parameter $\gamma$: At saturation, $\gamma = 3.2$, $R = 0.23 \mu\text{M}$ (top), the nucleation events are frequent but do not scan the entire DNA. At low RecA concentration, $\gamma = 0.32$, $R = 0.0023 \mu\text{M}$ (bottom), sparse nucleation events scan the whole DNA. In the strong fluctuation regime, $\gamma = 1$, $R = 0.023 \mu\text{M}$ (middle), frequent nucleation-disassembly events scan the entire DNA.

*Measurements of ATP-enhanced DNA discrimination.*—RecA binding to short ssDNA oligomers was measured by the fluorescence anisotropy (FA) signal of a tag attached to the 3' end of the ssDNA [9,14]. The signal couples to the rotational diffusion of the tag: A freely rotating tag will slow down upon RecA binding and hence will increase the FA signal. Ensemble average binding curves as a function of RecA concentration were taken at steady state in the presence of ATP (energy source ON) or its nonhydrolyzable analog ATPγS (energy source OFF). The effect of ATP on sequence and length discrimination is most significant close to the onset of binding where ssDNA coverage is partial. With FA it is possible to probe this onset at the nanomolar range, which is less accessible to conventional biochemical techniques. Since RecA binding is directional and the ssDNAs are short, a single contiguous RecA filament binds close to the 3' end. Calibration experiments done with poly-Thymine ssDNAs (data not shown) suggest that the 3'-end tag "counts" the first bound monomers close to tag, while a 5'-end tag senses the last monomers that fill up the ssDNA.

The present work shows that discrimination of ssDNA sequences is enhanced by utilizing ATP. With ATPγS, RecA can discriminate between ssDNA sequences provided they fold into significantly different secondary structures that present a barrier for binding [14]. We therefore chose nearly identical periodic ssDNAs having essentially no stable folds: $(TAC)_N = TACTAC...$ and $(TCA)_N$, of lengths $N = 13, 26$ (39, 78 bases) [15]. Binding with ATPγS is tight [Figs. 2(a) and 2(b)], exhibiting a sharp cooperative onset at low RecA concentration ($R_{onset} = 20$ nanomolar) and up to saturation we cannot discriminate between $(TAC)_{13}$, $(TAC)_{26}$, $(TCA)_{26}$, and $(TCA)_{13}$ (the latter not shown). Replacing ATPγS with ATP shifts the binding onset to higher RecA concentration ($R_{onset} = 70$ nanomolar). Now, with the energy source ON, the binding curves separate, exhibiting discrimination of length and sequence: Longer ssDNAs are favored over short ones [Fig. 2(a)], and $(TAC)_N$ over $(TCA)_N$ [Fig. 2(b); $N = 13$ is not shown]. Kinetics of binding upon a rapid increase of RecA concentration shows the transient to steady state with a time scale that increases with length [Fig. 2(c)]. The onset shift is expected due to the ATP-driven disassembly which destabilizes the bound RecA filament; a higher RecA concentration is required for stable binding. Less expected is the ssDNA discrimination induced by ATP and a theoretical explanation is proposed below.

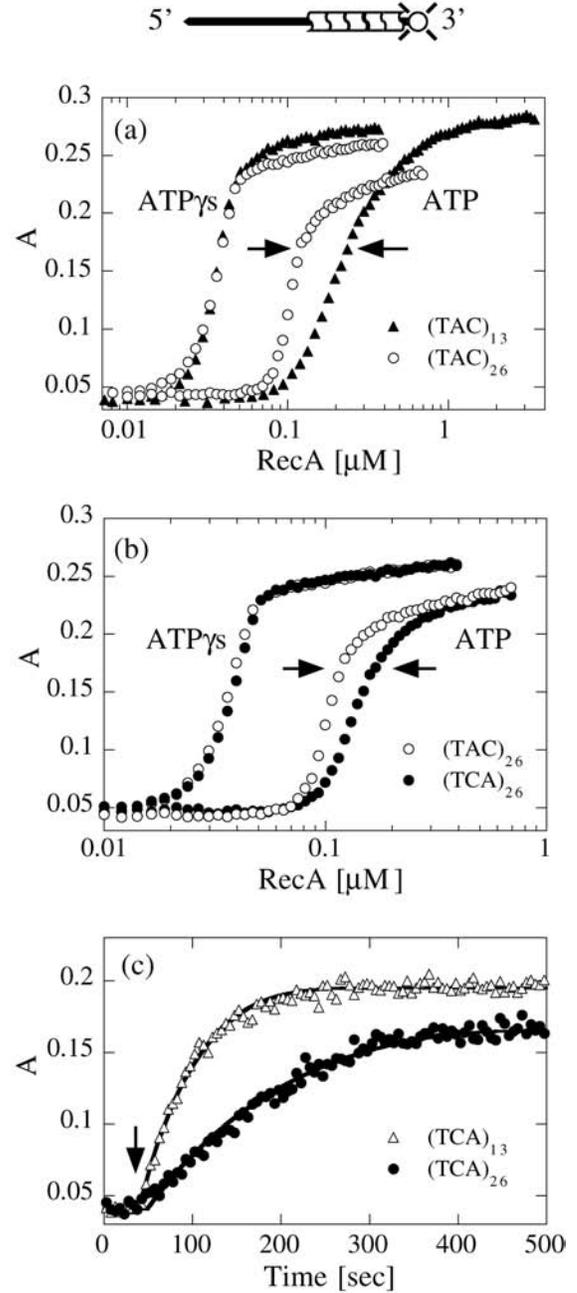

FIG. 2. RecA binding to ssDNA measured by fluorescence anisotropy $A$ of fluorophore at the ssDNA 3' end (cartoon). (a),(b) With ATP$_{\gamma s}$, binding to all four sequences, $(TAC)_{13}$, $(TAC)_{26}$, and $(TCA)_{13}$ (not shown) cannot be discriminated. With ATP, lengths (a) and sequences (b) can be discriminated. (c) Binding kinetics upon rapid increase of RecA concentration (arrow). Data fitted to Eq. (5) (solid line) with the rates $\kappa_- \simeq 0.06$ sec$^{-1}$ and $\kappa_{nuc} \simeq 0.002$ sec$^{-1}$ $\mu$M$^{-1}$. The polymerization rate can be estimated by $\kappa_-/R_{onset} \simeq 0.75$ sec$^{-1}$ $\mu$M$^{-1} \gg \kappa_{nuc}$, consistent with our model assumption.

*Kinetic proofreading and RecA binding cascade.*—We suggest that the enhanced DNA discrimination by energy-driven assembly is reminiscent of kinetic proofreading (KPR). To clarify, consider first an extension of the "classical" two-stage KPR to an $N$-stage pathway [Fig. 3(a)]. Initial reactants $Q_0$ progress *irreversibly* to final products $Q_N$ through a series of $N$ intermediates. For each intermediate $Q_n$, the reaction can either move forward to $Q_{n+1}$ at rate $\kappa_-$, or return to state $Q_0$ at rate $\kappa_+$. Backward reactions $Q_{n+1} \to Q_n$ are disfavored by coupling to an energy-driven process (ATP hydrolysis). At steady state, the influx at any intermediate stage is equal to the outflux $\kappa_- Q_n = \kappa_+ Q_{n+1} + \kappa_- Q_{n+1}$. It follows that the concentrations of intermediates decay exponentially $Q_n \sim K^{-n}$ ($K = (\kappa_- + \kappa_+)/\kappa_- > 1$ is the Michaelis constant). This leads to exponentially amplified discrimination between two competing pathways, $S$ and $G$. If pathway $S$ is disfavored with respect to pathway $G$, $K_G < K_S$, then the overall cascade discrimination, defined by the ratio of products, is $Q_N^S/Q_N^G \sim f^N$, where $f = K^G/K^S$ [16].

A similar cascade is constructed by the pathway of RecA assembly cascade [Fig. 3(b)]: It starts at stage $Q_0$ (a fully covered ssDNA) and moves forward through irreversible disassembly steps of the last, 5' end, monomer of the RecA filament. The intermediates $Q_n$ are RecA filaments of length $N - n$ that either progress to stage $Q_{n+1}$ by a subsequent disassembly or return to any of the previous stages $Q_0, ..., Q_{n-1}$ by nucleation followed by rapid filament extension. The cascade differs from the multistage KPR since at stage $Q_n$ there are $n$ available nucleation sites and the nucleation rate is hence proportional to $n$, in contrast to KPR scheme where the backward reaction rates are constant. As shown below, this leads to a qualitatively improved discrimination factor $Q_N^S/Q_N^G \sim f^{N^2/2}$.

*Out-of-equilibrium dynamics.*—The functionality of the RecA system as a sensor of DNA length relies on the ATP energy source that drives it far from equilibrium. Modeling therefore requires accounting of the stochastic dynamics beyond mean-field rate equations [8]. A master equation is derived under the following assumptions: (1) Polymerization is extremely rapid and nucleation anywhere on the ssDNA is instantly extended to the 3' end. (2) The ssDNA is short enough such that there is a single contiguous RecA filament. In our simple stochastic model we consider ssDNAs of length $N$, the number of available RecA binding sites. (The length in nucleic bases is $3N$ since each RecA monomer binds to a base triplet.) There are $N$ RecA binding states with probability $p_n(t)$ for a ssDNA with $n$ vacancies at the 5' end (filament length $N - n$) at time $t$. With the total nucleation rate proportional to RecA concentration, $\kappa_+ = \kappa_{\text{nuc}} \cdot R$ the master equation is ($0 \leq n < N$)

$$\frac{dp_n(t)}{dt} = -\kappa_-[p_n(t) - p_{n-1}(t)] + \kappa_+\left[\sum_{m=n+1}^{N} p_m(t) - np_n(t)\right], \quad (2)$$

with the boundary condition, ($n = N$) $dp_N(t)/dt = \kappa_- p_{N-1}(t) - \kappa_+ N p_N(t)$.

Expressed in terms of the cumulative probability $P_n = \sum_{m=n}^{N} p_m$ the probability that the filament is shorter than $N - n$, the master equation becomes $dP_n(t)/dt = -\kappa_-[P_n(t) - P_{n-1}(t)] - \kappa_+ n P_n(t)$. A continuous approximation is therefore

$$\frac{\partial P(n, t)}{\partial t} = -\kappa_- \frac{\partial P(n, t)}{\partial n} - \kappa_+ n P(n, t), \quad (3)$$

with the boundary condition $P(0, t) = 1$ [9]. The steady-state solution of Eq. (3) is Gaussian

$$P_s(n) = e^{-(\kappa_+/2\kappa_-)n^2}, \quad (4)$$

with the filament distribution $p_s(n) = (\kappa_+/\kappa_-) \times n e^{-(\kappa_+/2\kappa_-)n^2}$. In analogy to multistage KPR, the last reaction stage $Q_N = p(N)$, a naked ssDNA, shows exponential sensitivity to the rates and to RecA concentration, but the linear $n$ dependence of the total nucleation rate leads to the unusual Gaussian dependence on length [Eq. (4)]. The discrimination is tuned by the control parameter $\gamma = N^2 \kappa_{\text{nuc}} R/2\kappa_-$, as deduced by our previous scaling arguments [Eq. (1)].

*Mean-field and fluctuations.*—To appreciate the significance of fluctuations we show below that a mean-field approximation deviates from the stochastic dynamics.

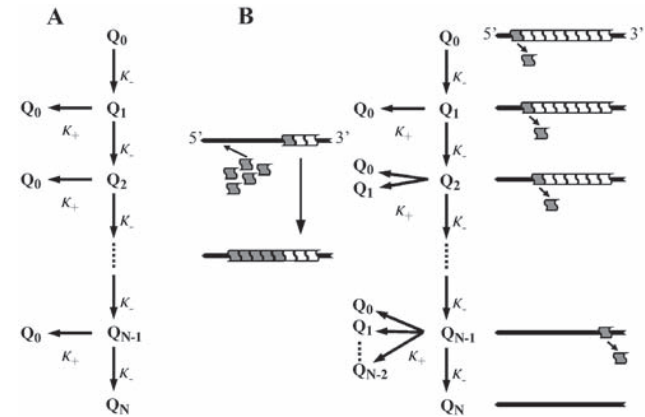

FIG. 3. Cascade architecture: (a) Generic multistage kinetic proofreading. (b) An analogous RecA assembly cascade (see text).

Summing over Eq. (2) we find the "hydrodynamic" relation $d\langle n\rangle/dt = \kappa_-[(1-p_N(t)] - \frac{1}{2}\kappa_+(\langle n\rangle^2 + \langle\Delta n^2\rangle)$, where the average (hole) occupancy is $\langle n\rangle = \sum n p_n(t)$ and the variance $\langle\Delta n^2\rangle = \sum(n-\langle n\rangle)^2 p_n(t)$. Within a mean-field approximation, one neglects the fluctuations and the amount of empty ssDNAs $\langle\Delta n^2\rangle = p_N(t) = 0$. The resulting mean-field equation is identical to the previously derived end-dependent disassembly kinetics [8], $d\langle n\rangle/dt = \kappa_- - \frac{1}{2}\kappa_+\langle n\rangle^2$. The steady-state mean-field occupancy is therefore $\langle n\rangle_{MF} = N/\sqrt{\gamma}$. We compare this result to the moments of the Gaussian steady-state solution ([Eq. (4)] $\langle n\rangle = N\sqrt{\pi}/(2\sqrt{\gamma})\mathrm{erf}\sqrt{\gamma}$ and $\langle\Delta n^2\rangle = N^2/\gamma[1 - e^{-\gamma} - \frac{\pi}{4}(\mathrm{erf}\sqrt{\gamma})^2]$. The "working point" of maximal sensitivity to DNA length is when $\gamma \gtrsim 1$ and the system is *strongly fluctuating*, $\langle\Delta n^2\rangle \sim \langle n\rangle^2$. In this regime, the mean-field average occupancy $\langle n\rangle_{MF}$ deviates significantly from the full stochastic average $\langle n\rangle$. Since the typical nucleation step is of length $N$, the fluctuations do not decay with DNA length in the thermodynamic limit. Hence, although the steady-state equation captures the averaged kinetics of RecA assembly, we need to consider the fluctuations to describe DNA length discrimination by kinetic proofreading.

An independent test of the present model is by comparison to kinetic measurements upon a sudden increase of RecA concentration [Fig. 2(c)]. The time dependent solution of Eq. (3), with the initial condition of empty ssDNAs $P(n,0) = 1$ is

$$P(n,t) = \begin{cases} P_s(n)e^{(\kappa_+/2\kappa_-)(n-\kappa_-t)^2} & :\kappa_- t \leq n, \\ P_s(n) & :\kappa_- t \geq n. \end{cases} \quad (5)$$

This solution describes the invasion of the DNA by the Gaussian steady-state profile $P_s(n)$ [Eq. (4)] behind a "shock wave" that moves at a constant speed $\kappa_-$ in the 5'-to-3' direction.

*An SOS trigger by RecA proofreading?*—Upon sudden DNA damage in the cell RecA rapidly binds to the exposed ssDNA gaps and "proofreads" their lengths. As a result, long ssDNAs are exponentially favored over short ones and an effective *binding transition* would occur for lengths above $N_c \sim \sqrt{2\kappa_-/R\kappa_{\mathrm{nuc}}}$, where $R$ is the cellular RecA concentration. The length dependence of RecA assembly fluctuations serves as a nonlinear switch. The *in vivo* RecA concentration ranges between 1–10 $\mu$M ($10^3$–$10^4$ proteins per cell). Assuming that the *in vitro* rates are not considerably different than in the cell, our measurement and model imply that the SOS response would be triggered when the exposed ssDNA gaps reach a critical length of $3N_c \simeq 10$–30 bases (RecA binds to base triplets). Thus, the high-fidelity RecA trigger can direct the SOS repair proteins towards the longer, more critical, damages.

We hypothesize that the observed RecA proofreading cascade may also be employed for pairing and homology search during recombination. Measuring the strong fluctuations predicted here on a single molecule requires significant reduction of RecA concentration [17]. The sensitivity of RecA to features of the DNA resembles predictions for DNA unzipping [18]. Recently, it has been shown that a similar cascade architecture enhances the recognition in the immune system [19]. This suggests that the multistep cascade is a possible cooperative design principle in noisy biological environment where enhanced fidelity is advantageous.

---